# Tuning the Magnetic Ordering Temperature of Hexagonal Ferrites by Structural Distortion Control

Kishan Sinha[1], Haohan Wang[1], Xiao Wang[2], Liying Zhou[3,4], Yuewei Yin[1], Wenbin Wang[5], Xuemei Cheng[2], David J Keavney[6], Huibo Cao[7], Yaohua Liu[7], Xifan Wu[3], Xiaoshan Xu[1,8]

[1]Department of Physics and Astronomy, University of Nebraska, Lincoln, Nebraska 68588, USA,
[2]Department of Physics, Bryn Mawr College, Bryn Mawr, Pennsylvania 19010, USA
[3]Department of Physics, Temple University, Philadelphia, Pennsylvania 19122, USA
[4]International Center for Quantum Materials and School of Physics, Peking University, Beijing 100871, People's Republic of China
[5]Department of Physics, Fudan University, Shanghai 200433, People's Republic of China
[6]Advanced Photon Source, Argonne National Laboratory, Argonne, Illinois 60439, USA
[7]Neutron Scattering Division, Oak Ridge National Laboratory, Oak Ridge, TN 37831, USA
[8]Nebraska Center for Materials and Nanoscience, University of Nebraska, Lincoln, Nebraska 68588, USA

## Abstract

To tune the magnetic properties of hexagonal ferrites, a family of magnetoelectric multiferroic materials, by atomic-scale structural engineering, we studied the effect of structural distortion on the magnetic ordering temperature ($T_N$). Using the symmetry analysis, we show that unlike most antiferromagnetic rare-earth transition-metal perovskites, a larger structural distortion leads to a higher $T_N$ in hexagonal ferrites and manganites, because the $K_3$ structural distortion induces the three-dimensional magnetic ordering, which is forbidden in the undistorted structure by symmetry. We also revealed a near-linear relation between $T_N$ and the tolerance factor and a power-law relation between $T_N$ and the $K_3$ distortion amplitude. Following the analysis, a record-high $T_N$ (185 K) among hexagonal ferrites was predicted in hexagonal ScFeO$_3$ and experimentally verified in epitaxially stabilized films. These results add to the paradigm of spin-lattice coupling in antiferromagnetic oxides and suggests further tunability of hexagonal ferrites if more lattice distortion can be achieved.

Spin-lattice couplings have a significant impact on magnetic properties. In antiferromagnetic (AFM) orthorhombic $RTM$O$_3$ (o-$RTM$O$_3$) for example, where $R$ stands for rare earth, Y, or Sc, and $TM$ stands for transition metal, a larger orthorhombic distortion from the cubic perovskite structure correlates with a lower Neel temperature ($T_N$) [see supplementary information], which may be understood as the reduction of the AFM super-exchange interactions caused by the smaller $TM$-O-$TM$ bond angles due to the orthorhombic distortions [1,2].

The effect of spin-lattice couplings may be employed to tune the magnetic properties. Here we focus on increasing the $T_N$ of hexagonal $R$FeO$_3$ (h-$R$FeO$_3$), a family of multiferroics materials that are promising candidates for applications because of their spontaneous electric and magnetic polarizations, and potential magnetoelectric effects due to the coupling between the ferroelectric and the magnetic orders [3,4]. For widespread applications, it is important to increase the $T_N$ of h-$R$FeO$_3$ [5], by, e.g. atomic-scale structural engineering based on the spin-lattice couplings.

On the other hand, in h-$R$FeO$_3$, $T_N$ increases with the lattice distortion, which is a puzzling trend opposite to that in the AFM o-$RTM$O$_3$ [see supplementary information]. Previously, Disseler et al. discovered a correlation between $T_N$ and lattice constants in h-$R$MnO$_3$ and h-$R$FeO$_3$ [6]. The higher $T_N$ for smaller $R$ has been attributed to closer Fe-Fe (or Mn-Mn) distances [6,7]. This understanding is worth revisiting, since it cannot explain that in AFM o-$RTM$O$_3$, the smaller lattice constants do bring the $TM$ atoms closer, but the reduced $TM$-O-$TM$ bond angles actually decreases the AFM exchange interactions and $T_N$. Hence, there should be a distinct mechanism of magnetic ordering and spin-lattice coupling in h-$R$FeO$_3$. Elucidating this mechanism will not only provide guidance in increasing $T_N$ of h-$R$FeO$_3$, but also add to the paradigms of spin-lattice coupling in AFM materials.

In this work, we examine the role of the structural distortion in the magnetic ordering in h-$R$MnO$_3$ and h-$R$FeO$_3$. A symmetry analysis shows that the three-dimensional magnetic ordering is forbidden in the undistorted structure by symmetry, but can be induced by the K$_3$ distortion with a power-law relation between $T_N$ and K$_3$ magnitude. Based on these revelations, we have predicted a record-high $T_N$ in h-$R$FeO$_3$ when $R$=Sc and experimentally confirmed it in epitaxially stabilized films.

Hexagonal ScFeO$_3$ (001) and YbFeO$_3$ (001) films (5 × 5 mm$^2$ and 10 × 10 mm$^2$ surface area, 70-200 nm thick) have been grown on Al$_2$O$_3$ (001) and yttrium stabilized zirconia (YSZ) (111) respectively using pulsed laser (248 nm) deposition in a 5 mTorr oxygen environment, at 750 °C with a laser fluence of about 1.5 J/cm$^2$ and a repetition rate of 2 Hz [8]. The film growth was monitored using the reflection high-energy electron diffraction (RHEED). The structural and magnetic properties have been studied using x-ray diffraction and spectroscopy, magnetometry and neutron diffraction. X-ray diffraction experiments, including θ/2θ scan, φ scan, and reciprocal space mapping were carried out using a Rigaku D/Max-B diffractometer with Co-Kα radiation (1.793 Å wave length) and a Rigaku SmartLab diffractometer with Cu-Kα radiation (1.5406 Å). X-ray absorption spectroscopy (including x-ray linear dichroism) with a 20° incident angle was studied at beamline 4-ID-C at the Advanced Photon Source at Argonne National Laboratory. Neutron diffraction experiments were carried out at beamline CORELLI at the Spallation Neutron Source (SNS) and HB3A four-circle diffractometer (FCD) at the High Flux Reactor (HFIR) with a thermal neutron wavelength of 1.546 Å, in the Oak Ridge National Laboratory. Temperature and magnetic-field dependence of the magnetization was measured using a superconducting quantum interference device (SQUID) magnetometer with the field along the film normal direction.

The crystal structure of isomorphic hexagonal $R$MnO$_3$ and $R$FeO$_3$ (h-$R$MnO$_3$ and h-$R$FeO$_3$) has a P6$_3$cm symmetry, consisting of alternating FeO (or MnO) and $R$O$_2$ layers [Fig. 1(a)]. AFM orders occur in h-$R$MnO$_3$ and h-$R$FeO$_3$ below about 70-140 K with spins in the FeO (or MnO) layers forming 120-degree structures [6,9–12]. Below about 1000 K, ferroelectricity in h-$R$MnO$_3$ and h-$R$FeO$_3$ is induced by a lattice distortion (K$_3$) [Fig. 1(a)] that tilts the FeO$_5$ (or MnO$_5$) local environment, shifts the $R$ atoms along the $c$ axis, and trimerizes the unit cell, with a sizable electric polarization ($P \sim 10$ μC/cm$^2$) [13–16]. In addition, hexagonal $R$FeO$_3$ exhibits a weak ferromagnetism [10,12,14,15,17,18] [Fig. 1(a)] due to the canting Fe spins.

Magnetic ordering relies on the underlying exchange interactions. In h-$R$FeO$_3$ and h-$R$MnO$_3$, although the exchange interactions within the FeO (or MnO) layers are strong, the inter-layer exchange interactions are weakened by the topology of layered structure and hexagonal stacking. Using h-$R$FeO$_3$ as an example, Fig. 1(b) shows the arrangement of the Fe atoms and their spins in two neighboring FeO layers. The Fe atoms are on the hexagonal A and C sites in the two layers respectively. One Fe atom (Fe$_0$) in the $z = c/2$ layer is highlighted by its tilted FeO$_5$ trigonal bipyramid. The interlayer nearest-neighbor exchange energy for Fe$_0$ is $E_{inter} = \sum_{i=1}^{3} J_{0i}\vec{S}_0 \cdot \vec{S}_i$, where $\vec{S}_i$ is the spin on Fe$_i$, and $J_{0i}$ is the exchange interaction coefficient between Fe$_0$ and Fe$_i$. When there is no lattice distortion, the local symmetry of Fe$_0$ is C$_{3v}$, leading to $J_{01}$=$J_{02}$=$J_{03}$ and $E_{inter}$=0 because $\sum_{i=1}^{3} \vec{S}_i = 0$. In other words, the interlayer exchange are canceled; the spin alignment between the two layers is lost. Therefore, the three-dimensional magnetic ordering is forbidden in the undistorted P6$_3$/mmc structure by symmetry.

On the other hand, the K$_3$ lattice distortion [Fig. 1(a)] reduces the symmetry to C$_S$, making $J_{01}$=$J_{02}$≠$J_{03}$. Consequently, nonzero lattice distortion leads to the three-dimensional magnetic ordering because $E_{inter}$=($J_{01}$-$J_{03}$)$S(S+1)$≠0 [19]. Since the inter-layer exchange interaction is the bottleneck of the three-dimensional magnetic ordering, one has $T_N \propto E_{inter}$= ($J_{01}$-$J_{03}$)$S(S+1)$. The dependence of $T_N$ on the K$_3$ distortion then hinges on the relation between $J_{01}$-$J_{03}$ and the magnitude of K$_3$ ($Q_{K3}$). Previously, Das et al. analyzed the relation between $J_{01}$-$J_{03}$ and $Q_{K3}$ [3]. Expanding $J_{01}$ and $J_{03}$ with respect to $Q_{K3}$ around $Q_{K3}$=0, the odd terms are expected to be zero due to the symmetry at $Q_{K3}$=0, leaving $J_{01}$-$J_{03} \propto a_2 Q^2_{K3} + a_4 Q^4_{K3}$, where a$_2$ and a$_4$ are coefficients. In Fig. 1(c), we plot the log$\{T_N/[S(S+1)]\}$ as a function of log($Q_{K3}$) of h-$R$MnO$_3$ measured using the neutron diffraction from the literature [20,21] [see supplementary information], where spin $S$ is 2 for Mn. The data appear to fall on a straight line, indicating a power-law relation $T_N/[S(S+1)] \propto Q^n_{K3}$; a fit shows n = 2.7±0.05. Given that the tilt of FeO$_5$ and MnO$_5$ caused by the K$_3$ distortion is on the order of 5 degrees [9,20,21] which is not so small, both the a$_2 Q^2_{K3}$ and the a$_4 Q^4_{K3}$ terms could play a role, resulting 2<n<4.

It is challenging to predict the $T_N$ using the direct dependence of $T_N$ on the K$_3$ distortion though, because the K$_3$ distortion, which involves the displacement of oxygen atoms, is difficult to measure precisely. Tolerance factor $t=(r_R+r_O)/(r_{TM}+r_O)\sqrt{2}$ where $r_R$, $r_{TM}$, and $r_O$ are atomic radius of $R$, $TM$ and oxygen, is a good measure of lattice distortion from the cubic perovskite structure in o-$RTM$O$_3$. It could also be used to gauge the structural distortion in h-$R$FeO$_3$ and h-$R$MnO$_3$, because a smaller $R$ atom is expected to reduce the in-plane lattice constant, which needs to be accommodated by a larger K$_3$ distortion to reduce the distances between Fe (or Mn) atoms within the FeO (or MnO) layers. In other words, smaller $t$ is expected to lead to larger $T_N$, which is consistent with the data from the literature [Fig. 1(c) inset] [11,20,21], where a linear correlation between $T_N/[S(S+1)$ and $t$ in h-$R$FeO$_3$ and h-$R$MnO$_3$ is observed ($S = 2$ and 2.5 for Mn and Fe respectively).

Following the trend in Fig. 1(c), a smaller $R$, corresponding to a smaller $t$, will lead to a higher $T_N$ in h-$R$FeO$_3$ and h-$R$MnO$_3$. Since Sc has much smaller atomic radius than the rare earth and Y [22], $T_N$ in h-ScFeO$_3$ is expected to be higher than that of other h-$R$FeO$_3$. To verify the prediction, we studied the magnetic ordering temperature in h-ScFeO$_3$. ScFeO$_3$ naturally crystallizes in bixbyite structure in bulk; high pressure growth of ScFeO$_3$ results in a corundum structure [23]. Previous studies show that partially substituting Lu with Sc in LuFeO$_3$ may stabilize the hexagonal structure [6,7,24]. However, the stabilization of pure ScFeO$_3$ in the P6$_3$cm structure has never been reported. In this study, we have successfully grown h-ScFeO$_3$ epitaxial films on Al$_2$O$_3$ (001) substrates. The crystal structure and epitaxial relations of the h-ScFeO$_3$ films were characterized using x-ray diffraction. As shown in Fig. 2(a), the θ/2θ scan shows a typical pattern of the P6$_3$cm structure with the epitaxial relation: h-ScFeO$_3$ (001) ∥ Al$_2$O$_3$ (001). The φ scan [see supplementary information] demonstrates the six-fold rotation symmetry and the in-plane epitaxial relation: h-ScFeO$_3$ (100) ∥ Al$_2$O$_3$ (100). The RHEED patterns [Fig. 2(b) and (c)], which are signatures of the h-$R$FeO$_3$ structure, indicate a flat surface. The FeO$_5$ trigonal bipyramid configuration is confirmed by the similarity between the x-ray linear dichroism spectroscopy of h-ScFeO$_3$ [Fig. 2(d)] and those of h-LuFeO$_3$ and h-YbFeO$_3$ observed previously [8,25–27]. From the x-ray reciprocal space mapping [see supplementary information], the lattice constants of h-ScFeO$_3$ were determined: $a$ = 5.742 Å and $c$ = 11.690 Å, smaller than the values of other h-$R$FeO$_3$ [28–30], suggesting a larger lattice distortion [31].

$T_N$ of h-ScFeO$_3$ was measured by the neutron diffraction experiments at CORELLI in addition to that of h-YbFeO$_3$. Using a wide wavelength-band neutron beam and a two-dimensional detector at CORELLI, a three-dimensional portion of the reciprocal space [using the Miller indices ($H$, $K$, $L$) as the coordinates] can be measured without rotating the sample [See supplementary materials]. The (101) and (114) diffraction peaks, were mapped out in the three-dimensional reciprocal space. As shown in Fig. 2(e), the two magnetic Bragg diffraction peaks (101) and (1-11), which are equivalent because of the six-fold rotational symmetry along the $c$ axis, were observed. The (101) Bragg peak is forbidden for the nuclear diffraction due to the crystal structure symmetry of h-$R$FeO$_3$ (space group P6$_3$cm), but it is allowed for magnetic diffraction [9]. The observation of the (101) peak confirms the magnetic ordering in h-$R$FeO$_3$, as previously shown in h-LuFeO$_3$ and h-$R$MnO$_3$ [6,9,12]. The temperature dependence of the (101) peak intensity suggests a transition at about 200 K, which is corroborated by the measurements at HB3A [Fig. 2(f)]. In contrast, the intensity of the (114) peak, which mainly comes from the nuclear scattering, shows an insignificant temperature dependence. As shown in Fig. 2(g), a similar transition temperature is observed in the temperature dependence of the magnetization measured using a SQUID magnetometer on warming, after cooling the sample in a 10 kOe magnetic field (field cool or FC) and after cooling in a zero magnetic field (zero-field cool or ZFC). The FC and ZFC curves diverge at around 185 K, giving a more precise determination of $T_N$.

As predicted, h-ScFeO$_3$ shows a high $T_N$ among all h-$R$MnO$_3$ and h-$R$FeO$_3$, as shown in Fig. 3(a), where our measurement on h-YbFeO$_3$ and data in the literature are also included [See supplementary information] [6,9,11–15,17,21]; the measured $T_N$ of h-ScFeO$_3$ is slightly lower than the value predicted by extrapolating the linear relation between $T_N$ and $t$, which is also true for h-ScMnO$_3$ [9]. The reduction of magnetization at low temperature in Fig. 2(g) hints a possible spin reorientation at about 100 K in h-ScFeO$_3$. However, the temperature dependence of the (101) peak intensity in Fig. 2(f) indicates that spin reorientation in h-ScFeO$_3$ may not be significant enough to change the spin structure from A$_2$ to A$_1$ [6].

Finally, we discuss the effect of lattice distortion on the canting of Fe moments, which is responsible for the net magnetization $M_{Fe}$ along the $c$ axis. The $M_{Fe}$ in h-ScFeO$_3$ can be inferred from the magnetometry data. As shown in Fig. 2(h), the $M$-$H$ curve shows a soft and a hard component, corresponding to two steps at $H \approx 0$ and $H \approx 30$ kOe respectively. This two-component feature has been observed in both h-LuFeO$_3$ and h-YbFeO$_3$ films [18,25]. The jump of magnetization at the higher field corresponds to the intrinsic coercivity of the h-$R$FeO$_3$, while the jump at low field corresponds to the unavoidable structural boundaries in film samples of h-$R$FeO$_3$ that create uncompensated spins. From the 30-kOe jump, we found that $M_{Fe}$ = 0.015+/-0.002 $\mu_B$/Fe in h-ScFeO$_3$, which is smaller than that of h-LuFeO$_3$ and h-YbFeO$_3$ [18,25], as shown in Fig. 3(b). This result is counter-intuitive, because a large K$_3$ distortion, corresponding to a larger tilt angle of the FeO$_5$ would seemingly generate a larger canting angle of the Fe moments ($\theta_{cant}$). However, $\theta_{cant}$ results from a competition between the exchange interaction and the Dzyaloshinskii-Moriya (DM) interaction [32–34]. Relation between the canting angle ($\theta_{cant}$) of the Fe moments, tilt angle of the FeO$_5$ ($\gamma_{tilt}$), lattice constant in the basal plane ($a$), and the intralayer exchange interaction coefficient $J$ can be derived as $\theta_{cant} \propto a^2 \gamma_{tilt}/J$ [See supplementary materials]. Although h-ScFeO$_3$ is expected to have a larger $\gamma_{tilt}$ and smaller $J$, $a$ is also smaller. Hence, the size of $\theta_{cant}$ cannot be simply linked to the amplitude of $\gamma_{tilt}$. If the effect of $a$ dominates, $M_{Fe}$ would decrease for smaller $R$, which is what we found in our previous first-principle calculations [31].

In conclusion, using symmetry analysis, we showed that the three-dimensional magnetic ordering in h-$R$MnO$_3$ and h-$R$FeO$_3$ are forbidden in undistorted structures by symmetry, but can be induced by the structural distortions. We also showed that dependence of $T_N$ on structural distortions manifests as a near-linear relation with the tolerance factor and a possible power law with $Q_{K3}$, suggesting a higher $T_N$ in h-ScFeO$_3$ with respect to other hexagonal ferrites studied so far, which was realized in this work in epitaxially stabilized films. In addition to indicating that the multiferroic ordering in h-$R$FeO$_3$ and h-$R$MnO$_3$ may be further enhanced with larger lattice distortions, these results also establish a paradigm of structural origin of magnetic ordering and spin-lattice coupling in AFM oxides.

Acknowledgement
This work was primarily supported by the National Science Foundation (NSF), Division of Materials Research (DMR) under Award No. DMR-1454618. X.W. was supported by National Science Foundation through Award No. DMR-1552287. X.M.C. acknowledges partial support from NSF DMR-1708790 and DMR-1053854. This research used resources at the Spallation Neutron Source and the High Flux Isotope Reactor, DOE Office of Science User Facilities operated by the Oak Ridge National Laboratory, and resources of the Advanced Photon Source, a U.S. Department of Energy (DOE) Office of Science User Facility operated for the DOE Office of Science by Argonne National Laboratory under Contract No. DE-AC02-06CH11357. The research was performed in part in the Nebraska Nanoscale Facility: National Nanotechnology Coordinated Infrastructure and the Nebraska Center for Materials and Nanoscience, which are supported by the National Science foundation under award ECCS: 1542182, and the Nebraska Research Initiative.

# References

[1] J. B. Goodenough, J. Phys. Chem. Solids **6**, 287 (1958).

[2] J. Kanamori, J. Phys. Chem. Solids **10**, 87 (1959).

[3] H. Das, A. L. Wysocki, Y. Geng, W. Wu, and C. J. Fennie, Nat. Commun. **5**, 2998 (2014).

[4] M. Ye and D. Vanderbilt, Phys. Rev. B **92**, 035107 (2015).

[5] J. A. Mundy, C. M. Brooks, M. E. Holtz, J. A. Moyer, H. Das, A. F. Rébola, J. T. Heron, J. D. Clarkson, S. M. Disseler, Z. Liu, A. Farhan, R. Held, R. Hovden, E. Padgett, Q. Mao, H. Paik, R. Misra, L. F. Kourkoutis, E. Arenholz, A. Scholl, J. A. Borchers, W. D. Ratcliff, R. Ramesh, C. J. Fennie, P. Schiffer, D. A. Muller, and D. G. Schlom, Nature **537**, 523 (2016).

[6] S. M. Disseler, X. Luo, B. Gao, Y. S. Oh, R. Hu, Y. Wang, D. Quintana, A. Zhang, Q. Huang, J. Lau, R. Paul, J. W. Lynn, S. W. Cheong, and W. Ratcliff, Phys. Rev. B **92**, 054435 (2015).

[7] A. Masuno, A. Ishimoto, C. Moriyoshi, N. Hayashi, H. Kawaji, Y. Kuroiwa, and H. Inoue, Inorg. Chem. **52**, 11889 (2013).

[8] S. Cao, X. Zhang, T. R. Paudel, K. Sinha, X. Wang, X. Jiang, W. Wang, S. Brutsche, J. Wang, P. J. Ryan, J.-W. Kim, X. Cheng, E. Y. Tsymbal, P. A. Dowben, and X. Xu, J. Phys. Condens. Matter **28**, 156001 (2016).

[9] A. Munoz, J. A. Alonso, M. J. Martinez-Lope, M. T. Casais, J. L. Martinez, and M. T. Fernandez-Diaz, Phys. Rev. B **62**, 9498 (2000).

[10] W. Wang, J. Zhao, W. W. Wang, Z. Gai, N. Balke, M. Chi, H. N. H. N. Lee, W. Tian, L. Zhu, X. Cheng, D. J. D. J. Keavney, J. Yi, T. Z. T. Z. Ward, P. C. P. C. P. C. Snijders, H. M. H. M. Christen, W. Wu, J. Shen, and X. Xu, Phys. Rev. Lett. **110**, 237601 (2013).

[11] M. Fiebig, T. Lottermoser, and R. V. Pisarev, J. Appl. Phys. **93**, 8194 (2003).

[12] S. M. Disseler, J. A. Borchers, C. M. Brooks, J. A. Mundy, J. A. Moyer, D. A. Hillsberry, E. L. Thies, D. A. Tenne, J. Heron, M. E. Holtz, J. D. Clarkson, G. M. Stiehl, P. Schiffer, D. A. Muller, D. G. Schlom, and W. D. Ratcliff, Phys. Rev. Lett. **114**, 217602 (2015).

[13] Y. K. Jeong, J. H. Lee, S. J. Ahn, and H. M. Jang, Chem. Mater. **24**, 2426 (2012).

[14] Y. K. Jeong, J. Lee, S. Ahn, S.-W. Song, H. M. Jang, H. Choi, and J. F. Scott, J. Am. Chem. Soc. **134**, 1450 (2012).

[15] S.-J. Ahn, J.-H. Lee, H. M. Jang, and Y. K. Jeong, J. Mater. Chem. C **2**, 4521 (2014).

[16] P. Murugavel, J. H. Lee, D. Lee, T. W. Noh, Y. Jo, M. H. Jung, Y. S. Oh, and K. H. Kim, Appl. Phys. Lett. **90**, 142902 (2007).

[17] H. Yokota, T. Nozue, S. Nakamura, H. Hojo, M. Fukunaga, P. E. Janolin, J. M. Kiat, and A. Fuwa, Phys. Rev. B **92**, 054101 (2015).

[18] J. A. Moyer, R. Misra, J. A. Mundy, C. M. Brooks, J. T. Heron, D. A. Muller, D. G. Schlom, and P. Schiffer, APL Mater. **2**, 012106 (2014).

[19] H. Wang, I. V. Solovyev, W. Wang, X. Wang, P. J. Ryan, D. J. Keavney, J.-W. Kim, T. Z. Ward, L. Zhu, J. Shen, X. M. Cheng, L. He, X. Xu, and X. Wu, Phys. Rev. B **90**, 014436 (2014).

[20] S. Lee, A. Pirogov, M. Kang, K.-H. H. Jang, M. Yonemura, T. Kamiyama, S.-W. S.-W. W.

Cheong, F. Gozzo, N. Shin, H. Kimura, Y. Noda, and J.-G. G. J.-G. Park, Nature **451**, 805 (2008).

[21] A. Munoz, J. Alonso, M. Martinez-Lopez, M. Cesais, J. Martinez, and M. Fernandez-Diaz, Chem. Mater. **13**, 1497 (2001).

[22] R. D. Shannon, Acta Cryst. **A32**, 751 (1976).

[23] M. Li, U. Adem, S. R. C. Mcmitchell, Z. Xu, C. I. Thomas, J. E. Warren, D. V Giap, H. Niu, X. Wan, R. G. Palgrave, F. Schiffmann, F. Cora, B. Slater, T. L. Burnett, M. G. Cain, A. M. Abakumov, G. Van Tendeloo, M. F. Thomas, M. J. Rosseinsky, and J. B. Claridge, J. Am. Chem. Soc. **134**, 3737 (2012).

[24] L. Lin, H. M. Zhang, M. F. Liu, S. Shen, S. Zhou, D. Li, X. Wang, Z. B. Yan, Z. D. Zhang, J. Zhao, S. Dong, and J.-M. Liu, Phys. Rev. B **93**, 075146 (2016).

[25] S. Cao, K. Sinha, X. X. Zhang, X. X. Zhang, X. Wang, Y. Yin, A. T. A. T. N'Diaye, J. Wang, D. J. D. J. Keavney, T. R. T. R. Paudel, Y. Liu, X. Cheng, E. Y. E. Y. Tsymbal, P. A. P. A. Dowben, and X. Xu, Phys. Rev. B **95**, 224428 (2017).

[26] W. Wang, H. Wang, X. Xu, L. Zhu, L. He, E. Wills, X. Cheng, D. J. Keavney, J. Shen, X. Wu, and X. Xu, Appl. Phys. Lett. **101**, 241907 (2012).

[27] S. Cao, X. Zhang, K. Sinha, W. Wang, J. Wang, P. A. Dowben, and X. Xu, Appl. Phys. Lett. **108**, 202903 (2016).

[28] X. Zhang, Y. Yin, S. Yang, Z. Yang, and X. Xu, J. Phys. Condens. Matter **29**, 164001 (2017).

[29] E. Magome, C. Moriyoshi, Y. Kuroiwa, A. Masuno, and H. Inoue, Jpn. J. Appl. Phys. **49**, 09ME06 (2010).

[30] A. A. Bossak, I. E. Graboy, O. Y. Gorbenko, A. R. Kaul, M. S. Kartavtseva, V. L. Svetchnikov, and H. W. Zandbergen, Chem. Mater. **16**, 1751 (2004).

[31] K. Sinha, Y. Zhang, X. Jiang, H. Wang, X. Wang, X. Zhang, P. J. Ryan, J.-W. Kim, J. Bowlan, D. A. Yarotski, Y. Li, A. D. DiChiara, X. Cheng, X. Wu, and X. Xu, Phys. Rev. B **95**, 094110 (2017).

[32] I. Dzyaloshinsky, J. Phys. Chem. Solids **4**, 241 (1958).

[33] T. Moriya, Phys. Rev. **120**, 91 (1960).

[34] F. Keffer, Phys. Rev. **126**, 896 (1962).

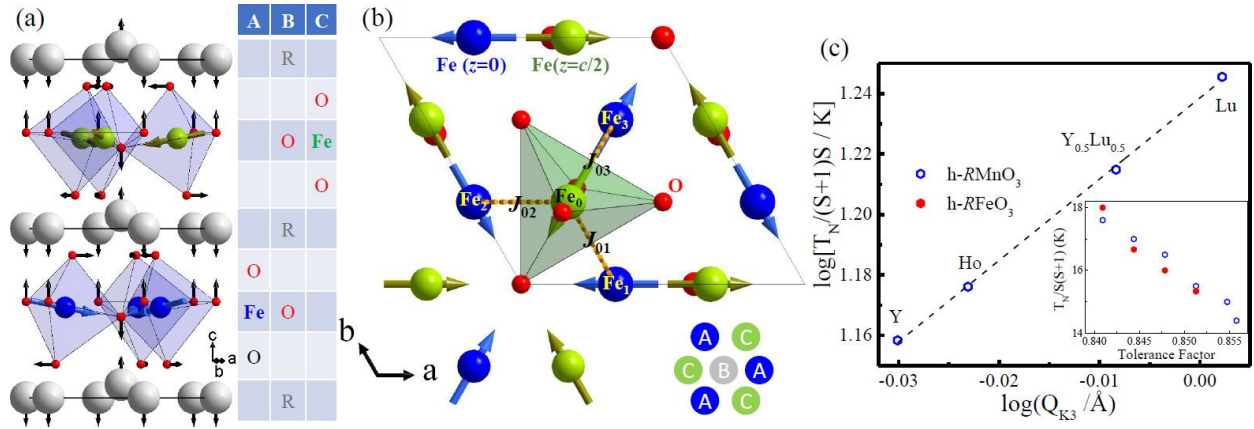

**Figure 1**. (a) Atomic structure of h-$R$FeO$_3$ depicted by a hexagonal unit cell. The arrows through the Fe atoms indicate the spins. The arrows from the atoms indicate the atomic displacements of the K$_3$ lattice distortion. The table indicates the hexagonal stacking. (b) The geometric arrangement of Fe atoms in the $z=0$ and $z=c/2$ layers. The arrows through the Fe atoms indicate the spin directions. The atom Fe$_0$ is highlighted by its FeO$_5$ trigonal bipyramid to depict the local symmetry. (c) $\log\{T_N/[S(S+1)]\}$ as a function of $\log(Q_{K3})$. Inset: $T_N/[S(S+1)]$ as a function of the tolerance factor. The dashed line is a linear fit to the data. The data are from the literature (see text).

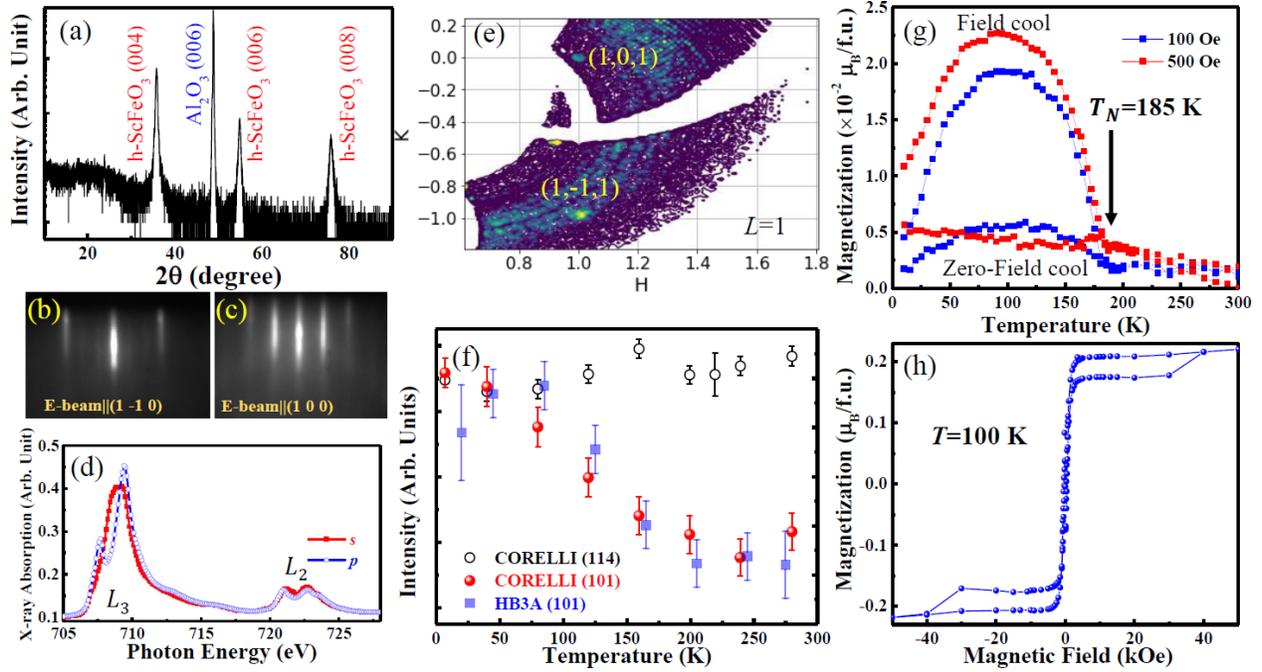

**Figure 2**. Structural and magnetic characterizations of h-ScFeO$_3$(001)/Al$_2$O$_3$ films. (a) θ/2θ x-ray diffraction measured using an x-ray wavelength 1.789 Å. (b) and (c) are the RHEED diffraction patterns measured when the electron beam are along the (1-10) and (100) directions. (d) X-ray absorption spectra measured at the Fe $L$ edges using $s$ (in plane) and $p$ (out of plane) linearly polarized x-ray beams. (e) A slice of the reciprocal space of h-ScFeO$_3$ at $L$=1 measured using neutron diffraction at CORELLI. (f) Temperature dependence of the neutron diffraction intensities of the (101) and (114) peaks measured at CORELLI and HB3A. (g) Temperature dependence of the magnetization per formula unit (f.u.) measured during warming after field-cool (10 kOe) and zero-field-cool using 100 Oe and 500 Oe. (h) Magnetization-field hysteresis loop measured at 100 K. The field is along the film normal direction.

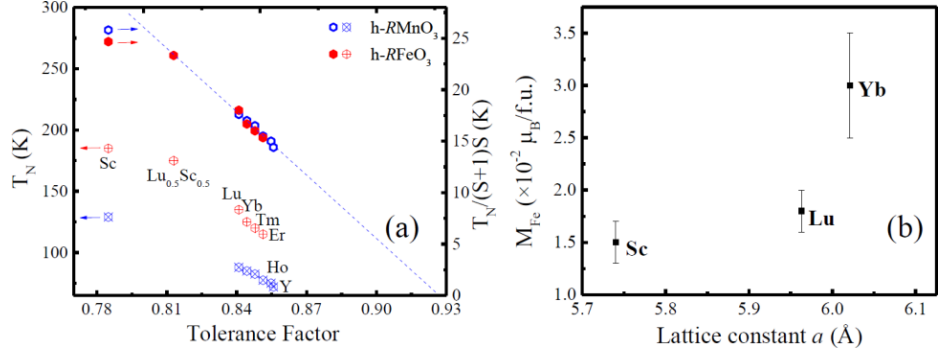

**Figure 3**. (a) The dependence of $T_N$ and $T_N/S(S+1)$ on the tolerance factor. The dashed line is a guide to the eye. (b) The magnetization from the canting of Fe spins ($M_{Fe}$) as a function of the in-plane lattice constant. Except for h-YbFeO$_3$ and h-ScFeO$_3$ in (a) and h-ScFeO$_3$ in (b), the data are from the literature (see text).

**Supplementary material**

S1. Neel temperatures of antiferromagnetic $RTM$O$_3$

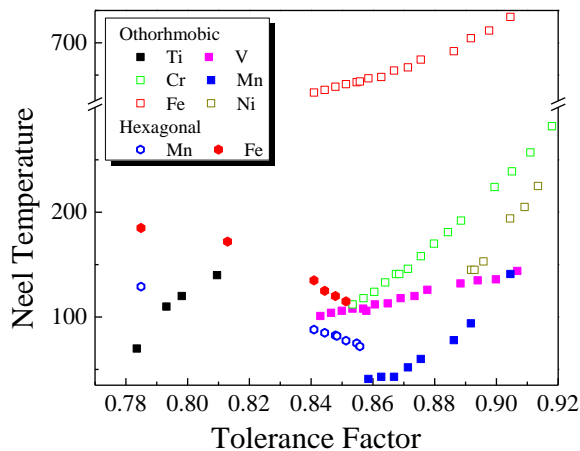

**Figure S1**. The Neel temperature of orthorhombic antiferromagnetic $RTM$O$_3$.

In Fig. S1, we surveyed the Neel temperature of antiferromagnetic $RTM$O$_3$ with orthorhombic and hexagonal structures, where $R$ stands for rare earth, Y, or Sc, $TM$ stands for transition metal. The results are plotted as a function of the tolerance factor [1,2], defined as $t=(r_R+r_O)/(r_{TM}+r_O)\sqrt{2}$, where $r_R$, $r_{TM}$, and $r_O$ are atomic radius of $R$, $TM$ and oxygen. For orthorhombic $RTM$O$_3$ (o-$RTM$O$_3$), when $t$ decreases from 1, corresponding to orthorhombic distortions from the cubic perovskite structure, $T_N$ decreases. In contrast, for hexagonal $RTM$O$_3$ (h-$RTM$O$_3$), when $t$ decreases, $T_N$ increases.

$R$TiO$_3$ crystalize in orthorhombic structure. When the size of $R$ is smaller than that of Sm, the materials become ferromagnetic [3].

$R$VO$_3$ and $R$CrO$_3$ both crystalize in orthorhombic structure; all members are antiferromagnetic [4–9].

$R$MnO$_3$ crystalize in orthorhombic structure when the size of $R$ is larger than that of Dy, with antiferromagnetic order [10–14]. Otherwise, $R$MnO$_3$ crystalize in hexagonal structure, with 120-degree antiferromagnetic order [15].

$R$FeO$_3$ crystalize in orthorhombic structure; all members exhibit antiferromagnetism [15]. The hexagonal structures can be stabilized using thin film epitaxy or doping, resulting the 120-degree antiferromagnetic order [16–21] similar to that in h-$R$MnO$_3$. The plot includes the new data from this work.

$R$NiO$_3$, except for LaNiO$_3$ which crystalizes in rhombohedra structure with a paramagnetism, have orthorhombic structures with antiferromagnetism [22–24].

## S2. K$_3$ lattice distortion in hexagonal $R$FeO$_3$ and $R$MnO$_3$

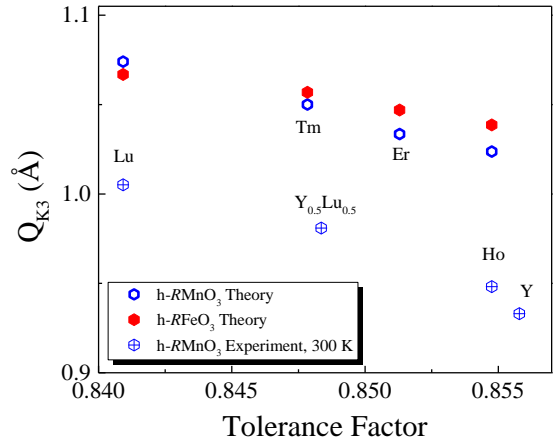

**Figure S2**. The K$_3$ lattice distortion in hexagonal $R$FeO$_3$ and $R$MnO$_3$.

In Fig. S2, we surveyed the K$_3$ lattice distortion of hexagonal $R$FeO$_3$ and $R$MnO$_3$. The experimental distortion amplitude ($Q_{K3}$) was calculated from the crystal structure measured using neutron diffraction, since the oxygen positions are critical [25,26]. The theoretical distortion amplitudes calculated based on the first principles [27,28], are systematically larger than the experimental values but follow the similar trend.

## S3. X-ray diffraction characterization of the h-ScFeO$_3$ films

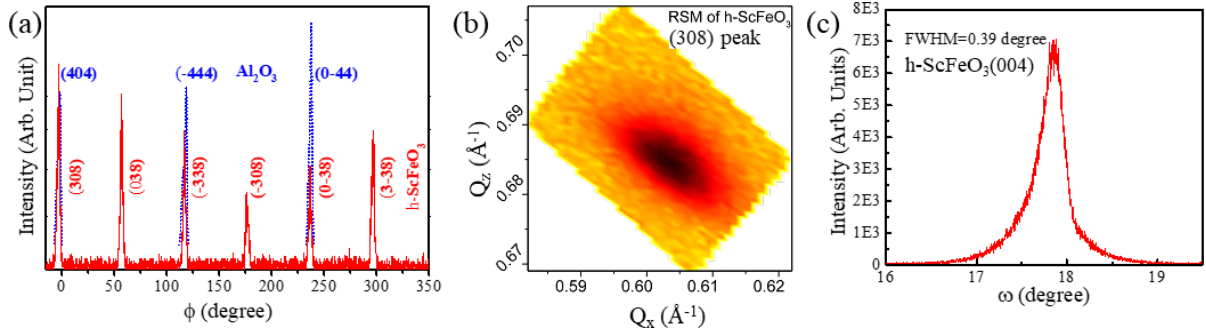

**Figure S3**. X-ray diffraction measurement on h-ScFeO$_3$(001)/Al$_2$O$_3$(001) films: (a) φ scan, (b) reciprocal space mapping (RSM) of the h-ScFeO$_3$ (038) peak, and (c) rocking curve on the h-ScFeO$_3$ (004) peak. The φ scan and RSM were measured on a Rigaku Smart Lab with 1.5406 Å x-ray wavelength while the rocking curve is measured using a Rigaku D/Max-B with 1.789 Å x-ray wavelength.

Figure S3 shows the x-ray diffraction characterization of h-ScFeO$_3$ films. Hexagonal coordinates are used for both the h-ScFeO$_3$ film and the Al$_2$O$_3$ substrate. As shown in Fig. S3(a), the three-fold and the six-fold rotational symmetries of Al$_2$O$_3$ and of h-ScFeO$_3$ respectively are revealed by the φ scan, consistent with their rhombohedral and hexagonal structures respectively. From the alignment of the peaks, the epitaxial relation in the basal plane h-ScFeO$_3$(100) || Al$_2$O$_3$(100) is revealed. In Fig. S3(b), the reciprocal space mapping around the h-ScFeO$_3$ (308) peak is plotted. From the peak position, one can calculate the lattice constants: $a$ = 5.742 Å, and $c$ = 11.690 Å. Figure S3(c) shows the rocking curve of the h-ScFeO$_3$ (004) peak. Using the Scherrer equation [30], one can estimate the in-plane size of the crystallites: $\frac{0.9\lambda}{\Delta\theta \cos\theta}$, where $\lambda$ is the x-ray wavelength, $\Delta\theta$ is the full width half maximum (FWHM) in Fig. S3(c); the result shows the in-plane crystallite size is approximately 25 nm.

## S4. Neutron diffraction measurements of the h-YbFeO$_3$ films

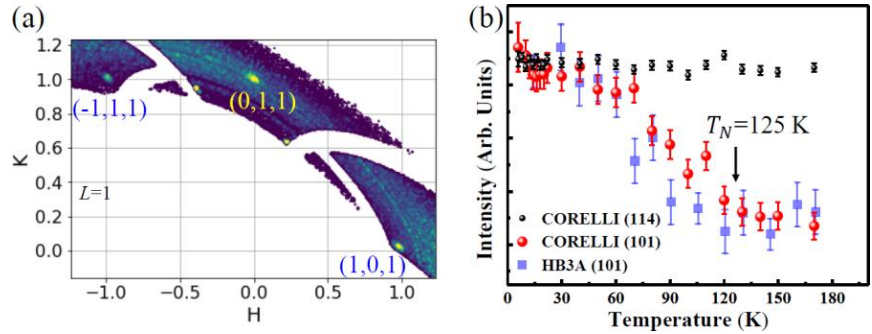

**Figure S4**. (a) A slice of the reciprocal space of h-YbFeO$_3$ at $L=1$ measured using neutron diffraction at CORELLI (see text). (b) Temperature dependence of the integrated neutron diffraction intensities of the (101) and (114) peaks measured at CORELLI and HB3A.

To study the magnetic ordering of h-YbFeO$_3$, we carried out neutron diffraction measurements. Figure S4(a) shows a slice of the reciprocal space at $L=1$. Three Bragg peaks (101), (011), and (-111), which are equivalent because of the six-fold rotational symmetry along the $c$ axis, are observed. The (101) Bragg peak is forbidden for the nuclear diffraction due to the crystal structure symmetry of h-$R$FeO$_3$ (space group P6$_3$cm), but it is allowed for magnetic diffraction [29]. Figure S4(b) shows the temperature dependence of the integrated peak intensities for selected peaks. A clear change of the slope of the (101) peak intensity is observed at $125 \pm 5$ K, suggesting a magnetic transition [Fig. S4(b)]. In contrast, the intensity of the (114) peak, which mainly comes from the nuclear scattering, shows an insignificant temperature dependence. The strong temperature dependence of the (101) peak is confirmed by single-crystal neutron diffraction at HB3A, as also shown in Fig. S4(b) and Fig. S5. The magnetic transition temperature of 125 K is consistent with the reported $T_N$ for h-YbFeO$_3$ from magnetometry [20].

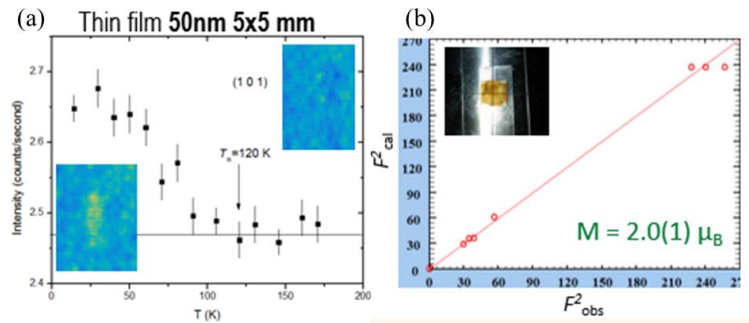

**Figure S5**. Neutron diffraction measurements on h-YbFeO$_3$ films at HB3A. (a) Diffraction intensity as a function of temperature for the (101) peak. The insets are the detector images showing the (101) peak at low temperature which vanishes at high temperature. (b) The calculated v.s. observed magnetic structural factors which lead to a magnetic moment of 2.0 $\mu_B$ per Fe.

## S5. Neutron diffraction data measured at the Spallation neutron source (SNS)

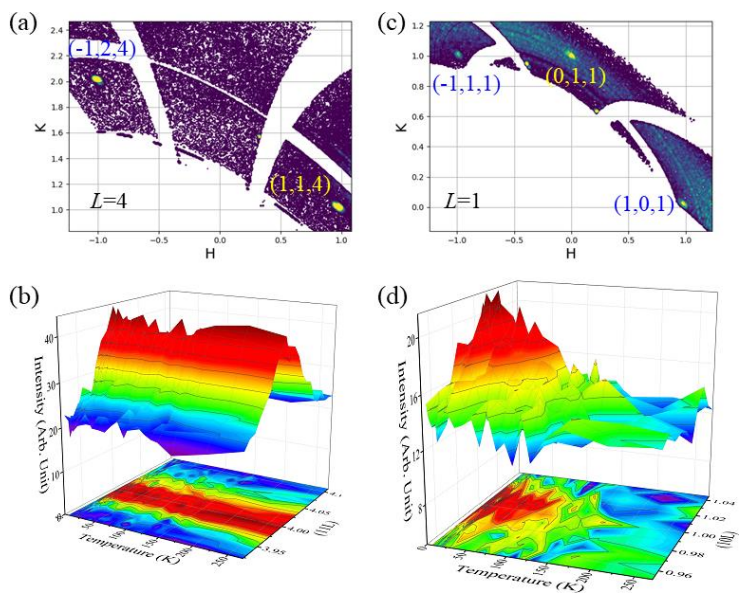

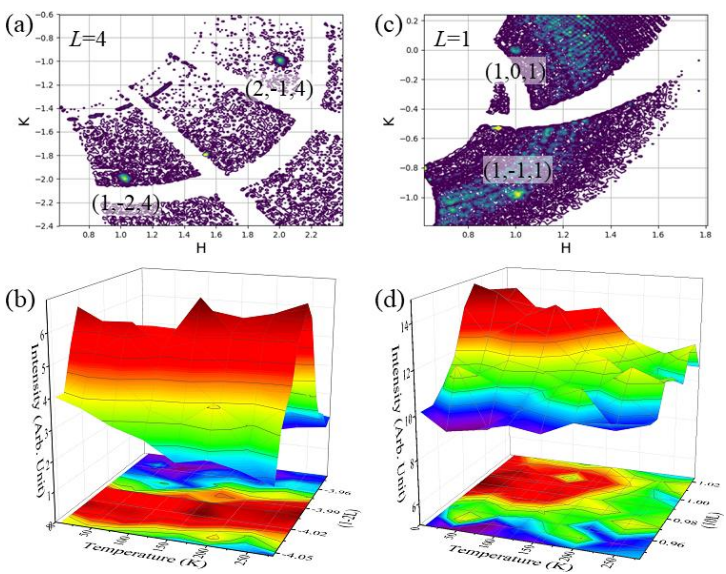

**Figure S6**. Neutron diffraction data of h-YbFeO$_3$. (a) and (c) are slices of three-dimensional reciprocal space at $L=4$ and $L=1$ respectively. (b) and (d) are the temperature dependence of the $L$ scan near the Bragg diffraction points (1-24) and (011) respectively.

**Figure S7**. Neutron diffraction data of h-ScFeO$_3$. (a) and (c) are slice of three-dimensional reciprocal space at $L=4$ and $L=1$ respectively. (b) and (d) are the temperature dependence of the $L$ scan near the Bragg diffraction points (1-24) and (101) respectively.

Elastic diffractions with unpolarized pulsed neutron beams were carried out on h-$R$FeO$_3$ films at the beamline CORELLI of the Spallation neutron source (SNS) with a pulsed neutron source. With the two-dimensional detector, and the pulsed neutron beam which contains neutrons with various wavelengths, a three-dimensional portion of the reciprocal space can be measured without rotating the sample. In this work, the incident angle of the neutron beam is set so that both the magnetic Bragg diffraction peak (101) and the nuclear Bragg diffraction peak (114) are included in this portion of the reciprocal space and measured at the same time.

Figure S6 and S7 show the raw data of the neutron diffraction on h-YbFeO$_3$ and h-ScFeO$_3$ respectively. Slices at $L=4$ and $L=1$ reveals the diffraction of (114) and (101) [Fig. S6 and S7 (a) and (b)] peaks and their equivalences respectively. The white space in the slices are parts of the reciprocal space that are not covered by the detector. From the temperature dependence of the *L* scan, one can observe the disappearance of the magnetic diffractions peak (101) but not the nuclear diffraction peak (114) within the temperature range of the measurements. It is obvious that the transition temperature of h-$ScFeO_3$ is significantly higher than that of h-$YbFeO_3$.

## S6. Dzyaloshinskii-Moriya (DM) interaction between Fe sites and the canting of the Fe spins in h-$R$FeO$_3$

- DM interactions

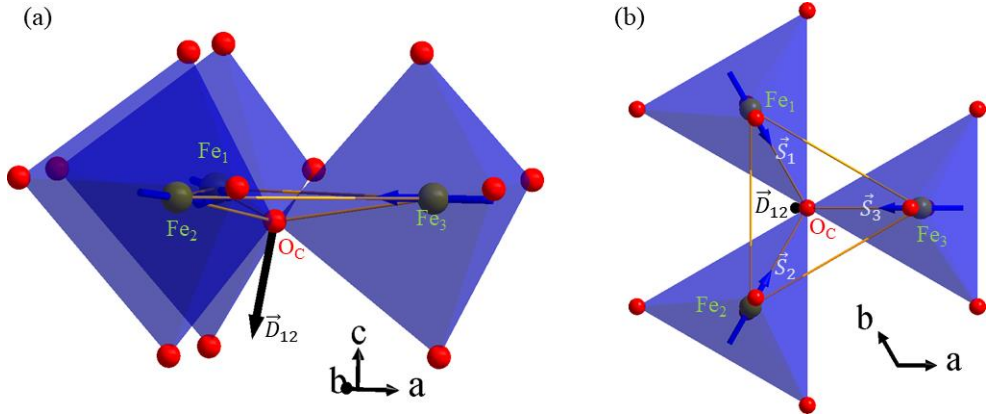

**Figure S8**. Schematics of the FeO$_5$ trimer. (a) Side view of the trimer highlighting the downward displacement of the atom O$_C$ and the DM interaction vector between Fe$_1$ and Fe$_2$. (b) Top view of the trimer.

Here we discuss the DM interaction [31,32] using the symmetry analysis and the Keffer's estimation [33]. As shown in Fig. S8, the Fe sites in h-$R$FeO$_3$ trimerizes due to the collective tilt of the FeO$_5$. We can assume the positions of the three Fe atoms (Fe$_1$ to Fe$_3$) and the oxygen in the center (O$_C$) as

$$\vec{r}_{Fe1} = \frac{a}{3}\left(-\frac{1}{2}, \frac{\sqrt{3}}{2}, 0\right)$$

$$\vec{r}_{Fe2} = \frac{a}{3}\left(-\frac{1}{2}, -\frac{\sqrt{3}}{2}, 0\right)$$

$$\vec{r}_{Fe3} = \frac{a}{3}(1,0,0)$$

$$\vec{r}_{O_C} = (0,0,-\delta),$$

where $a$ is the lattice constant in the basal plane of h-$R$FeO$_3$.

We analyze the DM interaction between Fe$_1$ and Fe$_2$ as an example. According to Moriya [31], the DM interaction between two magnetic ions results from the on-site spin-orbit coupling and the hopping between the electronic states of the magnetic ions through a diamagnetic atom (or atoms). The symmetry of the geometric arrangement of the two magnetic ions and the diamagnetic atom(s) is critical. For the DM interaction between Fe$_1$ and Fe$_2$, the symmetry of the Fe$_1$-O$_C$-Fe$_2$ connectivity determines the direction of the vector coefficient $\vec{D}_{12}$ in the DM interaction $\vec{D}_{12} \cdot (\vec{S}_1 \times \vec{S}_2)$. Two symmetry rules [31] can be applied here: 1) if there is a mirror plane including Fe$_1$ and Fe$_2$, $\vec{D}_{12} \perp$ the mirror plane; 2) if there is a mirror plane perpendicular to $\vec{r}_{Fe1} - \vec{r}_{Fe2}$ that passes the point $\frac{\vec{r}_{Fe1}+\vec{r}_{Fe2}}{2}$, $\vec{D}_{12} \parallel$ the mirror plane. Therefore, $\vec{D}_{12}$ is perpendicular to the plane including the Fe$_1$-O$_C$-Fe$_2$ atoms.

According to Keffer [33], if the DM interaction is mediated mainly by one diamagnetic atom, one can estimate the magnitude and the orientation of the vector interaction coefficient as $\vec{D}_{i,j} \propto \vec{r}_{i3} \times \vec{r}_{j3}$, where $\vec{r}_{i3}$ and $\vec{r}_{j3}$ are the vectors from the two magnetic ions to the diamagnetic (3$^{rd}$) atom respectively. In the case of Fe$_1$-O$_C$-Fe$_2$,

$$\vec{D}_{12} \propto \vec{r}_{Fe1O_C} \times \vec{r}_{Fe2O_C} = \left(\frac{a\delta}{\sqrt{3}}, 0, \frac{\sqrt{3}a^2}{18}\right),$$

where $\vec{r}_{Fe1O_C} = \vec{r}_{Fe1} - \vec{r}_{O_C}$ and $\vec{r}_{Fe2O_C} = \vec{r}_{Fe2} - \vec{r}_{O_C}$.

Assuming $\vec{D}_{12} = -D(\sin\phi, 0, \cos\phi)$, one finds

$$\tan\phi = \frac{6\delta}{a}.$$

Note that the tilt angle of the FeO$_5$ local environment can be defined using $\tan\gamma = \frac{\delta}{a/3} = \frac{3\delta}{a}$. Since all the angles are small (in h-LuFeO$_3$, $\gamma \approx 9.5$ degree), one finds approximately

$$\gamma \approx \frac{3\delta}{a}, \phi \approx \frac{6\delta}{a}.$$

In addition, since $\delta \ll a$, the magnitude of DM interaction strength follows

$$D \propto a^2.$$

- Canting of the Fe spin and the DM interaction

The canting of Fe spins mainly comes from the DM interaction and the tilt of the FeO$_5$ local environment which tilts the vector coefficient $\vec{D}_{i,j}$ (by angle $\phi$) and causes the Fe spins to reorient to minimize the DM interaction energy $\vec{D}_{i,j} \cdot (\vec{S}_i \times \vec{S}_j)$. At the same time, canting of the Fe spins reduces the Fe-Fe spin angle and increases the antiferromagnetic exchange interaction. Therefore, the canting angle $\theta$ is a result of the competition between the exchange interaction and the DM interaction. Below, we derive the relation between $\theta$ and $\phi$ (the tile angle of $\vec{D}_{i,j}$).

As shown in Fig. S8 we can assume the spin vectors as the following

$$\vec{S}_1 = \left(\frac{1}{2}\cos\theta, -\frac{\sqrt{3}}{2}\cos\theta, -\sin\theta\right)S$$

$$\vec{S}_2 = \left(\frac{1}{2}\cos\theta, \frac{\sqrt{3}}{2}\cos\theta, -\sin\theta\right)S$$

$$\vec{S}_3 = (-\cos\theta, 0, -\sin\theta)S,$$

where $S$ the magnitude of the Fe spin.

We consider the exchange interaction between the spins $\vec{S}_1$ and $\vec{S}_2$ as an example, which shows

$$E_{12}^{(ex)} = J\vec{S}_1 \cdot \vec{S}_2 = J\left(-\frac{1}{2} + \frac{3}{2}\sin^2\theta\right)S^2,$$

where $J > 0$ is the exchange interaction coefficient.

For the DM interaction, if we assume $\vec{D}_{12} = -D(\sin\phi, 0, \cos\phi)$, where $D > 0$ is the magnitude of the interaction (see above discussion), the interaction energy is

$$E_{12}^{(DM)} = \vec{D}_{12} \cdot (\vec{S}_1 \times \vec{S}_2) = -D(\sqrt{3}\sin\theta\cos\theta\sin\phi + \frac{\sqrt{3}}{2}\cos^2\theta\cos\phi).$$

The canting angle $\theta$ will be determined by the minimization of the total energy $E_{12}^{(ex)} + E_{12}^{(DM)}$. By noting that both $\theta$ and $\phi$ are small, one can make the approximation that

$$E_{12}^{(ex)} + E_{12}^{(DM)} \approx J\left(-\frac{1}{2} + \frac{3}{2}\theta^2\right) - D\left(\sqrt{3}\theta\phi + \frac{\sqrt{3}}{2}\right).$$

Therefore, the minimization results in

$$\theta \approx \frac{D}{\sqrt{3}J}\phi.$$

Since it is expected that the DM interaction is much weaker than the exchange interaction (or $D \ll J$), one expects $\theta \ll \phi$, which agrees with the experimental observation (typically $\phi$ is a few degrees while $\theta$ is a fraction of a degree).

According to the discussion above, $D \propto a^2$, where $a$ is the lattice constant of the basal plane. Hence the canting angle of the Fe spin follows

$$\theta \propto \frac{2a^2\phi}{\sqrt{3}J} = \frac{4a^2\gamma}{\sqrt{3}J}.$$

In the case of hexagonal ferrites, the smaller $R$ leads to larger $\phi$, which decreases $J$, but also decreases $a$. Thus, it is possible that the combined effect may lead to a reduction of the canting angle $\theta$ under the compressive strain.

[1] A. S. Bhalla, R. Guo, and R. Roy, Mater. Res. Innov. **4**, 3 (2000).

[2] R. D. Shannon, Acta Cryst. **A32**, 751 (1976).

[3] H. D. Zhou and J. B. Goodenough, J. Phys. Condens. Matter **17**, 7395 (2005).

[4] E. F. Bertaut, G. Bassi, G. Buisson, P. Burlet, J. Chappert, A. Delapalme, J. Mareschal, G. Roult, R. Aleonard, R. Pauthenet, and J. P. Rebouillat, J. Appl. Phys. **37**, 1038 (1966).

[5] T. Sakai, G. Y. Adachi, J. Shiokawa, and T. Shin-Ike, J. Appl. Phys. **48**, 379 (1977).

[6] A. Muñoz, A. Alonso, T. Casáis, J. Martínez-Lope, L. Martínez, and T. Fernández-Díaz, Phys. Rev. B - Condens. Matter Mater. Phys. **68**, 1 (2003).

[7] M. Reehuis, C. Ulrich, P. Pattison, B. Ouladdiaf, M. C. Rheinstädter, M. Ohl, L. P. Regnault, M. Miyasaka, Y. Tokura, and B. Keimer, Phys. Rev. B - Condens. Matter Mater. Phys. **73**, 1 (2006).

[8] M. Reehuis, C. Ulrich, K. Proke, S. Mat'A, J. Fujioka, S. Miyasaka, Y. Tokura, and B. Keimer, Phys. Rev. B - Condens. Matter Mater. Phys. **83**, 1 (2011).

[9] C. Ritter, S. A. Ivanov, G. V. Bazuev, and F. Fauth, Phys. Rev. B **93**, 1 (2016).

[10] F. Moussa, M. Hennion, J. Rodriguez-Carvajal, H. Moudden, L. Pinsard, and A. Revcolevschi, Phys. Rev. B **54**, 15149 (1996).

[11] Z. Jirák, J. Hejtmánek, E. Pollert, M. Maryško, M. Dlouhá, and S. Vratislav, J. Appl. Phys. **81**, 5790 (1997).

[12] A. Muñoz, J. A. Alonso, M. J. Martínez-Lope, J. L. García-Muñoz, and M. T. Fernández-Díaz, J. Phys. Condens. Matter **12**, 1361 (2000).

[13] J. G. Cheng, J. S. Zhou, J. B. Goodenough, Y. T. Su, Y. Sui, and Y. Ren, Phys. Rev. B **84**, 1 (2011).

[14] T. Kimura, G. Lawes, T. Goto, Y. Tokura, and A. P. Ramirez, Phys. Rev. B **71**, 224425 (2005).

[15] M. Fiebig, T. Lottermoser, and R. V. Pisarev, J. Appl. Phys. **93**, 8194 (2003).

[16] S. M. Disseler, J. A. Borchers, C. M. Brooks, J. A. Mundy, J. A. Moyer, D. A. Hillsberry, E. L. Thies, D. A. Tenne, J. Heron, M. E. Holtz, J. D. Clarkson, G. M. Stiehl, P. Schiffer, D. A. Muller, D. G. Schlom, and W. D. Ratcliff, Phys. Rev. Lett. **114**, 217602 (2015).

[17] H. Yokota, T. Nozue, S. Nakamura, H. Hojo, M. Fukunaga, P. E. Janolin, J. M. Kiat, and A. Fuwa, Phys. Rev. B **92**, 054101 (2015).

[18] S.-J. Ahn, J.-H. Lee, H. M. Jang, and Y. K. Jeong, J. Mater. Chem. C **2**, 4521 (2014).

[19] J. A. Moyer, R. Misra, J. A. Mundy, C. M. Brooks, J. T. Heron, D. A. Muller, D. G. Schlom, and P. Schiffer, APL Mater. **2**, 012106 (2014).

[20] Y. K. Jeong, J. Lee, S. Ahn, S.-W. Song, H. M. Jang, H. Choi, and J. F. Scott, J. Am. Chem. Soc. **134**, 1450 (2012).

[21] S. M. Disseler, X. Luo, B. Gao, Y. S. Oh, R. Hu, Y. Wang, D. Quintana, A. Zhang, Q. Huang, J. Lau, R. Paul, J. W. Lynn, S. W. Cheong, and W. Ratcliff, Phys. Rev. B **92**, 054435 (2015).

[22] M. L. Medarde, J. Phys. Condens. Matter **9**, 1679 (1997).

[23] J. A. Alonso, M. J. Martínez-Lope, M. T. Casais, J. L. Martínez, G. Demazeau, A. Largeteau, J. L. García-Muñoz, A. Muñoz, and M. T. Fernández-Díaz, Chem. Mater. **11**, 2463 (1999).

[24] M. T. Fernández-Díaz, J. A. Alonso, M. J. Martínez-Lope, M. T. Casais, and J. L. García-Muñoz, Phys. Rev. B **64**, 1444171 (2001).

[25] A. Munoz, J. Alonso, M. Martinez-Lope, M. Cesais, J. Martinez, and M. Fernandez-Diaz, Chem. … **13**, 1497 (2001).

[26] S. Lee, A. Pirogov, M. Kang, K. H. Jang, M. Yonemura, T. Kamiyama, S. W. Cheong, F. Gozzo, N. Shin, H. Kimura, Y. Noda, and J. G. Park, Nature **451**, 805 (2008).

[27] H. Tan, C. Xu, M. Li, S. Wang, B. L. Gu, and W. Duan, J. Phys. Condens. Matter **28**, 126002 (2016).

[28] C. Xu, Y. Yang, S. Wang, W. Duan, B. Gu, and L. Bellaiche, Phys. Rev. B **89**, 205122 (2014).

[29] A. Munoz, J. A. Alonso, M. J. Martinez-Lope, M. T. Casais, J. L. Martinez, and M. T. Fernandez-Diaz, Phys. Rev. B **62**, 9498 (2000).

[30] B. D. Cullity, *Elements of X-Ray Diffraction.* (Addison-Wesley Pub. Co., Reading, Mass., 1956).

[31] T. Moriya, Phys. Rev. **120**, 91 (1960).

[32] I. Dzyaloshinsky, J. Phys. Chem. Solids **4**, 241 (1958).

[33] F. Keffer, Phys. Rev. **126**, 896 (1962).